\documentclass[12pt,preprint]{aastex}

\usepackage{graphicx}
\usepackage{amsmath}
\usepackage{amssymb}
\usepackage{natbib}

\shorttitle{SAX J1810.8-2609}
\shortauthors{Zhu. et al}

\begin{document}

\title{Results from Long-Term Optical Monitoring of the Soft X-Ray Transient SAX J1810.8-2609}

\author{Ling Zhu\altaffilmark{1,}\altaffilmark{2}, Rosanne
  Di Stefano\altaffilmark{1}, Lukasz
  Wyrzykowski\altaffilmark{3,}\altaffilmark{4} }

\altaffiltext{1}{Harvard-Smithsonian Center for Astrophysics, 60
Garden St, Cambridge, MA 02138, USA}
\altaffiltext{2}{Department of Physics and Tsinghua Center for
Astrophysics, Tsinghua University, Beijing 100084, China;
zhul04@mails.tsinghua.edu.cn}
\altaffiltext{3}{Institute of Astronomy, University of Cambridge, Madingley Road, Cambridge CB3 0HA, UK}
\altaffiltext{4}{Warsaw University Astronomical Observatory, Al. Ujazdowskie 4, 00-478 Warszawa, Poland}

\begin{abstract}

In this paper we report the long-term optical observation of the faint
soft X-ray transient SAX J1810.8-2609 from OGLE and MOA. We have
focused on the 2007 outburst, and also did the cross-correlate between its
optical light curves and the quasi-simultaneous X-ray observations
from {\it RXTE/Swift}. 
Both the
optical and X-ray light curves of 2007 outburst show multi-peak features.
Quasi-simultaneous optical/X-ray luminosity
shows that both the X-ray reprocessing and viscously thermal emission
can explain the observed optical flux. There is a slightly X-ray delay
of $0.6\pm0.3$ days during the first peak, while the X-ray emission
lags the optical emission by $\sim 2$ days during the rebrightening
stage, which suggests that X-ray reprocessing emission contributes
significantly to the optical flux in the first peak, but the
viscously-heated-disk origin dominates the optical flux during
rebrightening. It implies variation of the physical environment of the
outer disk, even the source stayed in low-hard state during the whole
outburst.  The $\sim 2$-day X-ray lag indicates a small accretion
disk of the system, and the optical counterpart was not detected by
OGLE and MOA during quiescence, which constrained it to be fainter
than M$_I$ = 7.5 mag. There is a suspected short-time optical flare
detected at MJD = 52583.5 without X-ray counterpart detected, this
single flux increase may imply a magnetic loop reconnection in the
outer disk as proposed by Zurita et al (2003). The observations cover all stages of the outburst,
however, due to the low sensitivity of {\it RXTE}/ASM, we cannot
conclude whether it is an optical precursor at the initial rise of the
outburst.

\end{abstract}

\section{Introduction}

Neutron star low mass X-ray binaries (LMXB) were believed to behavior
similarly to the black hole LMXBs (Lin et al. 2007). They spend most
of the time in quiescence, and occasionally show an outburst with
accretion rate dramatically increased. During the outburst, the
neutron star LMXBs will evolve at the similar track of the state
transition as black hole LMXBs do, basically from a low/hard state to
a high/soft state (Remillard \& McClintock 2006).  The optical
emission of LMXBs was often contributed to thermal emission of the
companion star and outer accretion disk, and sometimes synchrotron
emission of a jet. The disk can be heated by mainly two physical
processes, one is the friction between adjacent layers of the disk
when the materials were accreted (Shakura \& Sunyaev 1973), the other
is the X-ray irradiation from the inner disk (van Paradijs \&
McClintock 1994).  With the evolution of disk structure and other
physical properties, the radiation mechanism of the optical flux will
vary, which will be different for neutron star LMXBs and black hole
LMXBs.

For black hole LMXBs, the synchrotron emission of a jet was believed
to dominate the optical emission during low/hard state, with possible
disk-reprocessed emission (Russell et al. 2006).  In the soft
state, all the near-infrared and some of the optical emissions are
suppressed, a behavior indicative of the jet switching off in
transition to the soft state (Russell et al. 2006). The viscously
heated disk emission may become the dominant source. The
multi-wavelength observation of GX 339-4 provided a perfect example to
demonstrate the trend: a good correlation among the fluxes from the X-ray
power-law component, the radio, and the optical was found during
low/hard state which suggests the jet emission, however, the optical
flux dramatically decreased when the source entered high/soft
state. Meanwhile, an $\sim$ two-week X-ray flux delay was found during
high/soft state which indicates a viscously heated disk emission.

For neutron star LMXBs, the jet emission is not important unless at
very high luminosity. X-ray reprocessing was believed to dominate the
optical emission at low/hard state, with possible contribution from
viscously heated disk (Russell et al. 2006). The long-time observation
of neutron star LMXB, e.g. Aquila X-1\footnote{It is noted that Aquila X-1 is a prototypical NS transient in many respects, but it is not so typical in terms of accretion disk.  It has a very long orbital period ($\sim19$ hr) and therefore a very large disk. Usually transient NSs have orbital periods in the 2-8 hr range. However, the quasi-simultaneous optical and X-ray  luminosity of Aquila X-1 follows the general relation for neutron star LMXBs (Russell et al. 2006)}, shows that neither the
optical/near-infrared color nor its brightness change sharply during
an X-ray spectral state transition. So it is believed that for Aquila
X-1 the outer accretion disk is not affected by X-ray spectral state
transitions (Maitra \& Bailyn 2008), the X-ray reprocessing was thought to
contribute most of optical emission at both the low/hard and high/soft
state.  
   
When the optical emission is dominated by the viscously-heated-disk
emission, the emission at each radius provides a measure of the
instantaneous local accretion rate at the given radius. The X-ray and
optical emission, respectively, map the mass flow through the inner
and outer disk. Continuous monitoring to both the X-ray and optical
emission allows us to track the temporal evolution of the system. The
cross-correlation of the X-ray and optical light curves helps to map
the accretion flow direction, the X-ray/optical time delay reflects
the viscous timescale of the disk (e.g. Homan et al. 2005).  The time
lag between the initial point of the outburst in X-ray and optical
emission was believed to be able to, to some extent, test the disk
model and the trigger mechanism of the outburst (Narayan et al. 1996).

SAX J1810.8-2609 is a soft X-ray transient (SXT) discovered on 10
March 1998 with the wide field cameras (2-28 keV) onboard the {\it
  BeppoSAX} satellite (Ubertini et al. 1998). It was identified as a
neutron star LMXB because a strong Type-I X-ray burst was detected
(Natalucci et al. 2000). The distance was estimated to be $\sim$4.9
kpc.

On 11-12 March 1998, a follow-up target of opportunity (ToO)
observation with the narrow field instrument onboard {\it BeppoSAX}
was performed with a total observing time of 85.1 ks. It showed a hard
X-ray spectrum with emission up to 200 keV. The broadband spectrum
(0.1 - 200 keV) can be described by two components: a soft black body
component with the temperature $T_{\rm BB} \sim$ 0.5 keV, and a
power-law component with the photon index $\Gamma = 1.96 \pm 0.04$
(Natalucci et al. 2000).

 From 1998 through 2007, SAX J1810.8-2609 had been in a quiescent
 state. The neutron star system in quiescence was also detected by
 {\it Chandra} on 16th August 2003 (Jonker et al. 2004). It
 had an unabsorbed X-ray luminosity of $\sim$10$^{32}$ erg s$^{-1}$
 over the energy range of 0.3-10 keV, given the distance of 4.9 kpc. It
 shows that the quiescent spectrum could be well fitted by the two
 models: 'neutron star atmosphere + power-law' model and 'black body +
 power-law' model. 
 
In August 2007, {\it Swift} detected a new phase of highly luminous
activity (Parson et al. 2007), and the luminosity varies between
(1.1-2.6) $\times10^{36}$ erg s$^{-1}$ during this
outburst. Considering the time interval of the recurrence, the
observed outburst luminosity corresponds to a low time-averaged
accretion rate of $5 \times 10^{-12}$ $\mathrm{M}_{\sun}$
$\mathrm{yr}^{-1}$ (Fiocchi et al. 2009).  The X-ray spectra had shown
the evolution during different epochs of the outburst, but a
significant power-law component was always present (Fiocchi et
al. 2009). It is noted that the source never reached the high soft
state during the outburst.

In this paper, we obtained the twelve-year optical light curve of SAX
J1810.8-2609, which covers two outbursts. We cross-correlated the
optical light curve with the X-ray light curve from the {\it RXTE}/ASM
and {\it Swift}/BAT archive. Section 2 describes the observations and
data calibration. In Section 3 we identify the optical counterpart and
analyze the temporal morphology of the outburst. In Section 4 we show
the results of the cross correlation and discuss their
implications. Section 5 is the summary.

\section{Observations}
\subsection{OGLE light curve}

The region containing SAX J1810.8-2609 has been regularly observed by
the Optical Gravitational Lensing Experiment (OGLE) project during
eleven observing seasons from 1997 to 2009, including the outbursts in
1998 March and 2007 August.

The OGLE data were collected using a dedicated 1.3m Warsaw telescope
located at Las Campanas Observatory, Chile, and operated by the
Carnegie Institution of Washington. During the years of 1996-2000 OGLE
used a single 2048$\times$2048 CCD chip (with a pixed size of 0.417\arcsec) operating in the drift-scan mode (OGLE-II; Udalski et
al. 1997).  For OGLE-III (2001-2009) the camera was upgraded to a
mosaic of eight $2048\times4096$ CCD chips with pixel size of 0.26\arcsec (Udalski 2003).  During OGLE-II and OGLE-III, for the Bulge
fields each observation was exposed for 120 seconds. The observations
were usually done in the I-band.

Datasets from both phases of OGLE were processed using the same
photometric package based on Difference Image Analysis (DIA; Wozniak
2000).  The method requires creation of a template/reference image,
composed of the best available images of a given field. The template
is then convolved with the PSF of an image to be analyzed and the
subtraction is performed, revealing residuals due to changes in the
objects' fluxes, or to the appearance of new objects. The flux
residuals are then measured with aperture photometry, and light curves
are created (Udalski 2008).

In templates for both OGLE-II and OGLE-III we found a faint star with
a constant $I\approx 19.5$ mag, 0.44\arcsec away from the outburst position as
shown in Figure 2.  Because there was no object at the outburst
position on the templates, all the residual flux from the outbursts
was attributed to this faint star.  Therefore the composite light
curve, shown in Figure 1, contains both the baseline fluxes from the template
object and outbursts, and the faint template object contributes a
negligible amount of emission during outbursts.

\subsection{MOA light curve}
This source is labeled in the Microlensing Observations in
Astrophysics (MOA) microlensing alert catalogue as
MOA\_2007\_BUL\_365\footnote{http://www.phys.canterbury.ac.nz/moa
  /microlensing\_alerts.html}(Bond et al. 2001). MOA was also
operating in the I-band. In quiescence, this source is also well below
the detection threshold of MOA. It was monitored during the 2007
outburst. We calibrated the MOA light curve using the
quasi-simultaneous OGLE light curve.

\subsection{X-ray monitoring data}
The All Sky Monitor (ASM) (2-12 keV) onboard the Rossi X-ray Timing
Explorer ({\it RXTE}) has provided the longest continuous coverage of
luminous X-ray sources (Levine et al. 1996). In its quiescent state,
SAX J1810.8-2609 was not detected by the ASM, but the ASM detected and
monitored both outbursts. The second outburst was also monitored by
{\it Swift} Burst Alert Telescope (BAT; $\sim$15-50 keV) through the
Hard X-ray Transient Monitor program (Barthelmy et al. 2005). We
retrieved both data sets from their respective
websites\footnote{http://xte.mit.edu/ASM\_lc.html} \footnote{http://heasarc.gsfc.nasa.gov/docs/Swift/results/transients/}. In
addition, {\it Integral}/IBIS (22-68 keV) also detected this source
during this outburst (Fiocchi et al. 2009), however, we did not
include the {\it Integral} data because the monitoring was not
frequent enough near peak.

\section{Identification and the outburst morphology}
\subsection{Identification}
In the optical band, the outburst is located in OGLE III field
BLG251.8 with the coordinates (R.A=18$^{h}$10$^m$44$^s$.47,
Dec.=-26$^{\circ}$09$^{'}$01$^{''}$.7) . This position is consistent
with the position of SAX J1810.8-2609 (R.A.=18$^{h}$10$^m$44$^s$.47,
Dec.=-26$^{\circ}$09$^{'}$01$^{''}$.2) determined by {\it Chandra}
(Jonker et al 2004), with an uncertainty of 0.6\arcsec. It is also
consistent with the optical counterpart position of
(R.A.=18$^{h}$10$^{m}$44$^{s}$.4 Dec.=-26$^{\circ}$09$^{'}$00$^{''}$)
with an uncertainty of 1\arcsec determined by Greiner et al
(1998). The non-detection by MOA and OGLE constrains the source to be
fainter than 21 mag in I band during quiescence (Figure 2).

In the optical light curve, we indicate the 1998 outburst with light
blue `$\diamond$' and the 2007 outburst with dark blue `$\ast$' (see
Figure 1).  It is noted that the light blue points are concentrated on
the decay phase in 1998 outburst because the main part was not
observed. 

The optical observations were usually done in the I-band, however, one
observation in R band were also available. We can use the magnitude
difference between I and R bands to estimate the disk temperature in
the outer region.  The R-band observation is R=$19.5\pm0.5$ on March
13th 1998 (Greiner et al. 1998), and two OGLE I-band observations
closest in time are: I=$18.5 \pm 0.1$ mag on March 10 and I=$18.7 \pm
0.1$ mag on March 16, respectively. So we take the average for the two
I-band detections, and use it as the proxy for the I-band magnitude on
March 13th. Therefore, we have R-I$=0.9 \pm 0.6$ mag. Once we make the
reduction correction (Sumi 2004), it is R-I$=0.35 \pm 0.6$ mag, and it
corresponds to a temperature $=3400\pm1000$ K with a black body model,
which is consistent with the typical temperature for the outer disk.

The 2007 August outburst was well monitored, and the whole optical and
X-ray light curves are shown in Figure 3.  The hard (15-50 keV) and
soft (2-10 keV) X-ray data are from {\it Swift}/BAT and {\it
  RXTE}/ASM, respectively. The MOA and OGLE data were combined
together to form the optical light curve. The similarity of the
optical and the X-ray light curve during 2007 outburst confirms the
identification of the optical counterpart.  We did not use the {\it
  Integral}/IBIS (22-68 keV) light curve because of the incomplete
covering.

In Figure 1, we notice that the fluxes at four occasions (which are
all out of the outburst phase) were significantly higher than the
baseline. It is certain that three of these are not likely to be real
because the other stars in the field also exhibited the similar flux
feature during these observations. It means that these flux
measurements were affected by the systematic uncertainties that may be
related to weather, sky background, or instrumental effects. The
fourth point with an elevated flux at MJD = 52583.5 is indicated with
a red circle. We don't have any direct evidence to show this point is
also affected, but the quality of the image makes it difficult for us
to conclude that the system was experiencing a true increase in
optical flux. If the flux increase is true, this data point is pretty
interesting, and it may represent an short-duration (no longer than 2
days) outburst. We checked all the dwell data of the corresponding {\it RXTE}/ASM
observations and found no X-ray detection. This single flux increase
may imply a magnetic loop reconnection in the outer disk as proposed
by Zurita et al (2003).

\subsection{The outburst morphology}

The 2007 August outburst showed a multi-peak morphology (see Figure
3). We divided the whole outburst light curve into three parts: (1) an
initial rise ($54314 < $ MJD $<54323$), where the outburst is detected
in optical but not in X-ray; (2) the first peak ($54323<$ MJD $<
54337$); and (3) the rebrightening stage ($54337<$ MJD $<54400$),
which includes all the subsequent peaks. The temporal behavior in the
optical and X-ray bands is very similar. The solid lines in Figure 3
is used to separate these three different stages. The position of
significant optical peaks/dips were indicated with dotted lines. The
corresponding X-ray peaks/dips usually follow the optical peaks/dips
but with a-few-days lag.  The first optical peak occurred at
$\mathrm{MJD} \sim 54330$, while the corresponding X-ray flux peaked
slightly later. The optical light curve has a significant dip when it
tends to get a second peak at $\mathrm{MJD} \sim 54346$, then a sudden
rise roughly three days later. The dip was followed by the X-ray light
curve about two days later. The third significant optical peak
occurred at $\mathrm{MJD} \sim 54360$, and was followed by both the
soft and hard X-ray light curve a few days later. The optical light
curve also has a wide peak near $\mathrm{MJD} \sim 54381$ which is
very similar to that in the hard X-ray light curve; the optical
emission still leads the X-ray emission.

\section{Cross analysis of the optical/X-ray data during the 2007 August outburst}
 
\subsection{The initial rise of the outburst}

The source was not detected by {\it RXTE}/ASM until the flux reached
the level of $\sim 10$ mCrab on MJD 54323 (Levine et al. 1996), but
the optical emission was detected as early as on MJD $\sim 54314$.
During the initial rise phase, the optical detection (OGLE and MOA)
was earlier than the X-ray detection ({\it RXTE}/ASM) by $\sim 9$
days.  Similar X-ray delays were found in quasi-simultaneous X-ray and
optical monitoring of SXT GRO J1655-40 (Orosz et al. 1997), XTE
J1550-564 (Jain et al. 2001), 4U 1543-47 (Buxton \& Bailyn 2004) and
Aquila X-1 (Shahbaz et al. 1998; Maitra \& Bailyn 2008), among which
the initial point of outburst in X-rays lagged in optical by $3-11$
days.  The X-ray delays at the initial rise reflect the timescale of
the thin disk, which determines the time it takes for the accretion
flow to arrive at the inner region where the X-ray emission comes
from.  It already takes into account the timescale by the material to
fill in the {\it Advection-Dominated Accretion Flow} (ADAF) region,
which can extend to large radius in quiescent state (e.g. Narayan \&
Yi 1995). For all the systems aforementioned, the initial X-ray rises
of the outbursts were all determined by linear extrapolations of the
{\it RXTE}/ASM light curve.

We notice, however, that the detection threshold for {\it RXTE}/ASM is
much higher than the flux of the source in quiescence state. The X-ray
flux might have increased by a factor of a few tens before detected.
Furthermore, the flux may actually increased in a way much slower than
linear increase before detected by {\it RXTE}/ASM, the initial point
found by linear extrapolation may be actually much later than
the actual starting point (Homan et al. 2005).  Here we apply a power-law
extrapolation to the X-ray/optical data points during the rise phase
of the outburst as shown in Figure 4.  It would make a big difference
if the flux increased exponentially rather than linearly. The X-ray
luminosity was calculated with a distance of 4.9 kpc (Natalucci et
al. 2000). Based on {\it Chandra} observations, its quiescent
luminosity is $\sim 10^{32}$~erg~s$^{-1}$ (Jonker et al. 2004). Figure
4 shows an extrapolation that indicates that optical flux may have begun to rise at MJD
   $54288^{+10}_{-10}$ with $23.2 > I > 21$ magnitude in quiescence. 
   The fit to the X-ray data has large uncertainty, it gave an X-ray initial rise at at MJD $\sim
   54293^{+12}_{-35}$. The start points of optical and X-ray flux are consistent with each other within one-sigma error. It is significantly different from the start points directly detected. We can not conclude X-rays delay at the
start of the outburst due to the limited data.

\subsection{Optical and X-ray cross correlation}
 
We conducted cross-correlations of the optical and X-ray emission for
the first peak and the rebrightening phases, respectively.  During the
first peak, we only cross-correlated the soft X-ray ({\it RXTE}, 2-12
keV) and optical light curves. However, during the rebrightening
stage, we cross-correlated optical and soft X-ray, optical and hard
X-ray ({\it Swift}, 15-50 keV), soft and hard X-ray emission,
respectively.

For the convenience of performing the cross-correlation, we did a
linear interpolation to each light curve with a time interval of 0.1
days. The cross-correlation between two interpolated light curves gave
out a {\it Cross Correlation Factor} (CCF) function with uniform time
lags. The factor CCF will tell us how strong the correlation is, and
it is defined as
$\mathrm{CCF}=\frac{1}{n-1}\sum{\frac{(f_i-f_0)(h_i-h_0)}{\sigma_f\sigma_h}}$,
where $\{f_i\}$ and $\{h_i\}$ represent the data series with finite
length, $f_0$ and $h_0$ are the averages of $\{f_i\}$ and $\{h_i\}$,
$\sigma_f$ and $\sigma_h$ are the variance of $f_i$ and $h_i$,
respectively. One caveat is that the value of CCF may not be zero even
if the two data series are not correlated at all, e.g., if the data
series are short, the noise may produce a non-zero CCF.

We therefore conducted a Monte Carlo (MC) simulation to create faked
X-ray and optical light curves using the same temporal interval and
uncertainty of each data point from the observed ones. In order to
make sure the faked light curves have no correlation, we assume each
value on the faked light curves is taken from a gaussian distribution
(its central value is set zero, and $\sigma$ of the distribution is
set to be same as the observed uncertainty at each epoch). Therefore,
the faked light curves generated in this way are like the pure
noise. Once one set of faked light curves are generated, we did a linear
interpolation and cross-correlated the interpolated light
curves, obtaining the time lag and the maximum CCF.

By repeating this process for 2000 times for the first peak phase and
the rebrightening phase, respectively, we computed the distribution of
the maximum values of CCF, and then derived the $3\sigma$ confidence
level of CCF of each phase. The $3\sigma$ confidence level for the
cross-correlation for the first peak is CCF$=0.70$, which is plotted
in Figure 5, and for the rebrightening phase it is CCF$=0.5$ as
plotted in Figure 6.

Any correlation with CCF above the limit value is likely to be due to
the real signal.  We employed the area centroid of the CCF function
above the limit value to derive the time lag (Koen. 2003). The
uncertainty of time lag was obtained by applying a MC simulation
described above, with considering the light curve uncertainties, to
compute the distribution of time lags. All the uncertainties in our
results are given but the $1\sigma$ level of confidence.

During the first peak (Figure 5), the soft X-ray and optical emission
are positively correlated with maximum CCF of 0.85, which has a 5.0
$\sigma$ level of confidence. The soft X-ray lags optical emission by
0.60$\pm$0.30 days.  During the rebrightening stage (Figure 6), the
soft and hard X-ray emission are correlated with maximum CCF of 0.61
(4.3$\sigma$ level of confidence), they vary simultaneously with no
obvious lag (time lag = $0.20\pm0.61$ days). The soft X-ray and
optical emission are weakly correlated with maximum CCF of 0.54
(3.5$\sigma$ level of confidence), soft X-ray lags optical emission by
2.10 $\pm$ 0.28 days. The hard X-ray and optical emission are
correlated with maximum CCF of 0.61(4.3$\sigma$ level of confidence),
hard X-ray lags optical emission by $1.80\pm 0.29$ days.
  

\subsection{The origin of optical emission}

The luminosities of the quasi-simultaneous optical and X-ray (2-10
keV) observations are shown in Figure 7. The optical-X-ray detection
pairs were used when (1) there is a X-ray detection within one day of
an optical detection, and (2) the confidence level for the X-ray
detection is larger than $1\sigma$. We adopted the approximation
L$_{\mathrm{opt}} \approx \nu {\mathrm F}_{\nu,\mathrm{I}}$ to
estimate the optical luminosity (we are approximating the spectral
range to the central wavelength of I band) the same as Russell et
al.(2006) did. The optical absorbed flux were dereddened using the
extinction map from Sumi (2004). Luminosities were calculated with the
distance $d=4.9$ kpc (Natalucci et al. 2000).

The disk will be heated by the friction between adjacent layers of the
disk when the materials were accreted (Shakura \& Sunyaev 1973), we
chose the relation $L_{\mathrm{opt}} \sim nL_{\mathrm{X}}^{0.5}$ for
the intrinsic thermal emission of a viscously heated disk (Frank et
al. 2002, Russell et al. 2006).  The disk can be also heated by the
X-ray and ultraviolet (UV) emission from the inner disk. In this case,
$L_{\mathrm{opt}} \sim nL_{\mathrm{X}}^{0.5}a$ was taken for optical
emission from X-ray reprocessing (van Paradijs \& McClintock 1994),
where $a$ is the orbital separation.  The flat optically-thick
spectrum of the jets of neutron star can also be extented to the
optical regime, we adopted the relation $L_{\mathrm{opt}}\sim
nL_{\mathrm{X}}^{1.4}$ (Migliari \& Fender 2006) for this process. The
theoretical relations for the viscously heated disk model and jet
model are plotted on Figure 7, and the normalization $n$ for each
model was taken from Russell et al (2007). The orbital separation of
SAX J 1810.8-2609 is not determined at present, we can not compare the
data with the X-ray reprocessing model directly. Russell et al (2006)
got the best fit of $L_{\mathrm{opt}}\sim nL_{\mathrm{X}}^{0.63}$ for
neutron star LMXBs in the hard state, most of which are consistent
with X-ray reprocessing model. The Russell-2006-best-fit line was
plotted on Figure 4 as an indicator of contribution from X-ray
reprocessing.

Furthermore, in Figure 7, we label the data points at different stages
using different symbols. During the initial rise stage, only the upper
limit of X-ray luminosity was available which gave no tight
constraint.  The luminosity ratios at the first peak and the
rebrightening stage are similar, while that at the rebrightening stage
is slightly higher.  We can see that the jet emission model can hardly
contribute such bright optical emission, while both the viscously
heated disk model and the Russell-2006-best-fit will suffice.

Considering the cross-correlation result, it is likely that the
optical emission is a combination of emissions with different
origins. The viscously-heated-disk emission will cause the X-ray
emission delay the optical emission by a viscous timescale. The X-ray
and optical emission will be simultaneous or the optical lags X-ray by a
few seconds for the X-ray reprocessing and jet mechanism. While doing
cross-correlation between light curve $L1$ and $L2$, if $L1$ is a
combination of two flux-nearly-equal light curves with different
lags ($La$ and $Lb$) to $L2$, it will give out a lag between $La$ and
$Lb$ (Zhu \& Zhang. 2010).

At the first peak, the $0.6\pm0.3$-day X-ray delay is not significant,
which suggests that the optical emission may be a combination of
emissions from the viscously heated disk and X-ray reprocessing. Each
component contributes a large portion of the optical light.  The X-ray
delay becomes significant, reaching to $\sim 2$ days at the
rebrightening stage, which suggests that viscously-heated-disk
emission becomes dominant. The prevail of viscously-heated-disk
emission should result from two sides: First, as the luminosity ratio
of the optical to the X-ray becomes slightly higher at rebrightening
stage, there should be an enhancement of radiation efficiency of
viscously heated disk. It was believed the outburst will have a rebrighteing stage if
 the disk was only partly irradiated, part of the disk was in shadow at the first peak (Truss et al. 2002). 
 The first peak is caused by the
accretion of gas within the irradiation portion of the disc, while the
subsequent peaks are caused by the accretion of the gas in the outer
disk, until the whole disk is kept in high-viscosity state. It is natural that the radiation area of the disk at
rebrightening stage will be larger, and the temperature may be higher.
Second, the X-ray reprocessing emission should be suppressed. It was
affected by the structure of both the X-ray source and the outer disk, if
the X-ray source is elevated above the disk or the outer disk is
warped, It will have high reprocessing efficiency. When the X-ray
source becomes aligned with the disk plane, the reprocessing efficiency
will decrease.  Here the change of X-ray delay between the first peak
and rebrightening stage is significant, the source should evolve outer
disk (maybe also inner disk) variation during the outburst.


\subsection{Extremely faint optical counterpart in quiescence}

The $\sim 2$ days X-ray delay during rebrightening stage properly
reflects the timescale of the disk during hard state. It also reflects
the outside-in disk accretion process, as we discuss in the outburst
morphology section, the variation of X-ray flux follows the optical
properly. Compared to the $\sim 18$ days X-ray/optical delay of GX
339-4 during rebrightening stage when the source has entered soft
state, the $\sim 2$ days X-ray delay is short. SAX J1810.8-2609 should
has a very small accretion disk. In addition, the source stayed in
hard state during rebrightening, it should distribute some time to
fill the ADAF region within the $2$ days.  An extremely small disk and
small companion star are expected, which is actually what people
expected for a faint SXT as they sorted SAX J1810.8-2609 (Heise et
al. 1999; King 2000).

In quiescence, we only have the detect limits for the optical flux
available: m$_I > 21$ mag and m$_R > 21.5$ mag (Greiner et al. 1998).
They correspond to M$_I > 7.5$ and M$_R > 8.0$ with d = 4.9 kpc
(Natalucci et al. 2000), respectively.  If the source follows the same
luminosity ratio during quiescent state and hard state, we need go
roughly 2.2 magnitude deeper in I-band to detect the optical
counterpart during quiescence with respect to the viscously heated
disk model. The optical counterpart in quiescent state is extremely
faint consistent with the very small disk.

\section{Summary and Conclusions}
With the OGLE and MOA light curves, we have identified the optical
counterpart to SAX J1810.8-2609 during outburst. Two outbursts,
occurring in 1998 and in 2007, were covered by OGLE, and the 2007
August outburst was monitored by both MOA and OGLE. Except the two
outbursts, there is a possibly short-time optical flare without X-ray
counterpart at MJD = 52583.5, which may imply a magnetic-loop
reconnection in the outer disk as proposed by Zurita et al (2003)

The initial X-ray rise of the 2007 outburst detected by {\it RXTE}/ASM
is $9$-day later than the optical detection by OGLE/MOA. However, the
flux-detection threshold of {\it RXTE}/ASM played a critically role to
this kind of X-ray delay, we have shown that an exponential
extrapolation is also reasonable for the data, which indicates no
X-ray delay at the initial rise. We conclude that no obvious X-ray
delay was detected at the initial rise with the present detection
threshold.

The optical light curve shows multi-peak, as does the X-ray light
curve during the 2007 outburst. The variation of X-ray follows the
optical flux properly, which maps the the outside-in disk accretion
process.  During the first peak phase, the soft X-ray emission lags
the optical emission by $0.6\pm0.3$ days, the X-ray reprocessing
should contribute significantly to the optical flux during the first
peak. There is an obvious $\sim 2$-day X-ray delay during the
rebrightening stage which suggests the viscously-heated-disk emission
becomes the dominate optical source. 
The X-ray reprocessing emission usually dominates the optical flux of the neutron star LMXB at low/hard state (Russell et al. 2006) which is probably the case for the first peak of SAX J1810.8-2609. The rebrightening stage is something unusual.
Unlike Aquila
X-1, the outer disk of SAX J1810.8-2609 should undergo physical
environment variations, the physical properties at inner disk may
also be different at the first peak and rebrightening stage, even the
source stayed in low/hard state during the outburst.  The $\sim 2$
days time lag implies a very small-radii disk and fairly small
companion star. The non-detection of the source by either OGLE or MOA
at the limit of $m_I = 21$ mag ($M_I = 7.5$ mag) is consistent with
this conclusion.

OGLE and MOA have been observing the Galactic bulge for over a
decade. OGLE has published a catalogue with $\sim 2\times10^5$
variable stars based on OGLE II data (Wozniak et al, 2002). The OGLE
III data will be available soon. These datasets may contain a large
number of potential rebrightening events due to LMXBs, and will provide
an unique opportunity for LMXBs study in the future.

 \acknowledgments We thank the anonymous referee for very helpful and
 constructive comments to improve the paper significantly. This work was supported in part by NSF under AST-0908878. We are
 grateful to Martin C. Smith for promoting this collaboration. L.Z.
 thanks Ian A. Bond for sharing the MOA data, Jeffrey E. McClintock and
 Lijun Gou for helpful discussion and suggestions.

\begin{figure}
 \centering\includegraphics[width=5.5in,height=3.4in]{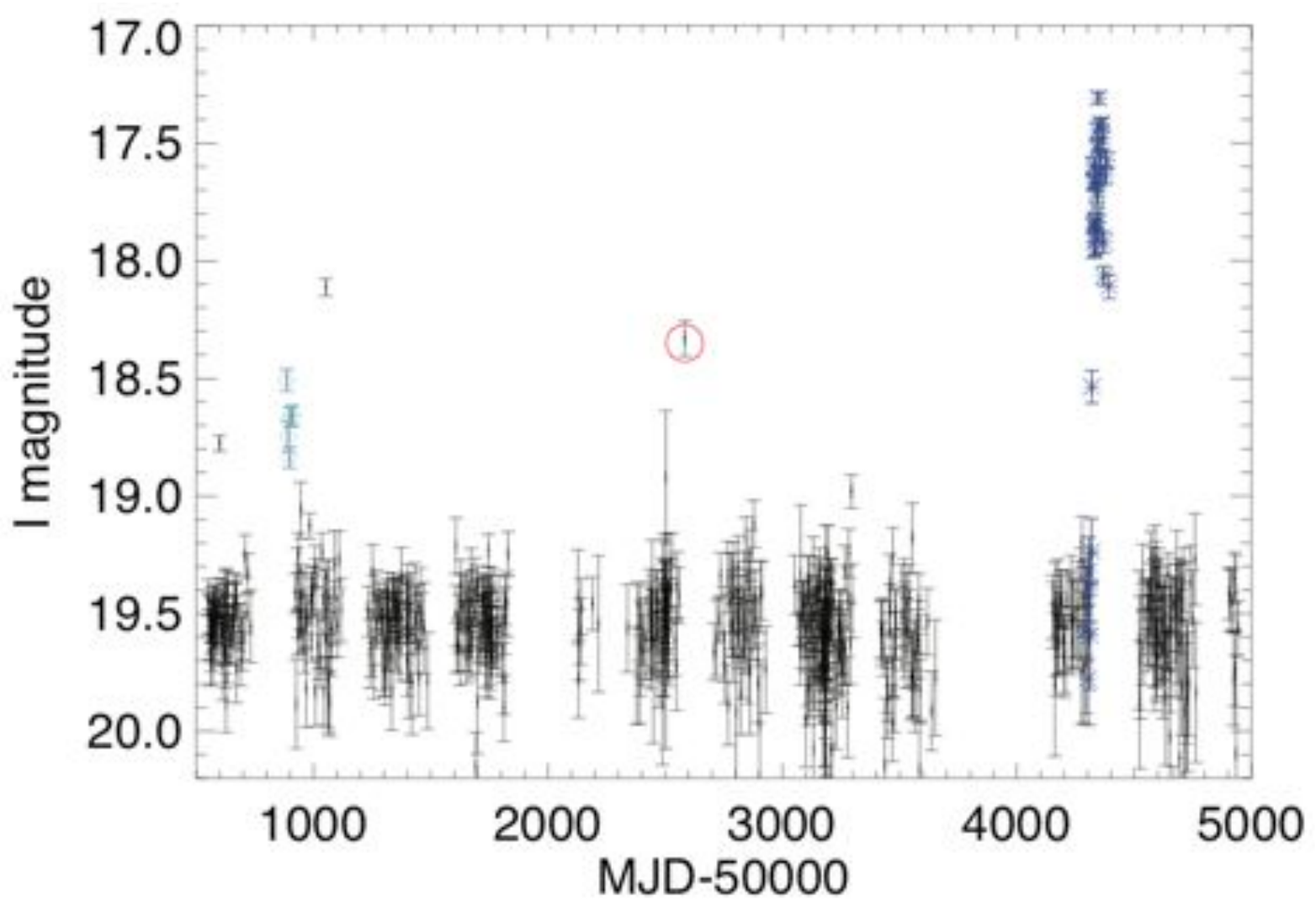}
 \caption{The optical light curve for the source SAX J1810.8-2609
   spanning over 12 years. The baseline is dominated by a faint nearby
   star. The light blue symbol `$\diamond$'s near MJD = 50800-50900
   and the dark blue symbol `$\ast$'s near MJD = 54300-54400 mark the
   outbursts at 1998 March and 2007 August, respectively. The OGLE
   team observed 4 higher-than-average-flux points outside the
   outburst phase, and among these only the measurement labeled with
   red circle has a good chance of representing a true higher
   flux. (See text.) }
 \label{Fig.1}
\end{figure}

 \begin{figure}
 \centering\includegraphics[width=5in,height=3in]{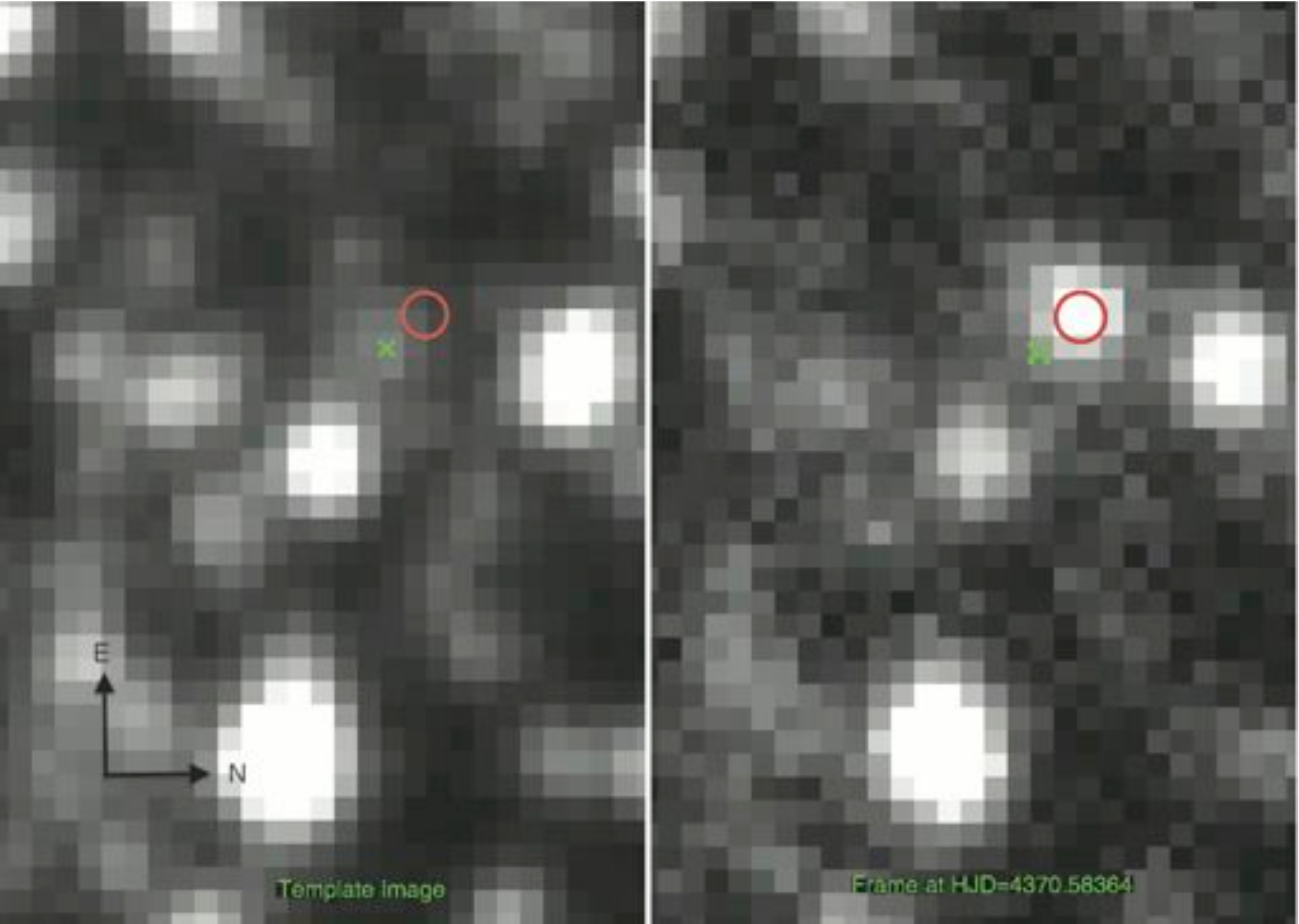}
 \caption{The image of the source during quiescence phase (left) and
   during outburst phase (right) from OGLE. The red circle marks the location of
   the outburst.  There is no detection during quiescence. The green
   `$\times$' represents the faint star nearby with a constant $I \approx 19.5$
   mag. }
 \label{Fig.2}
\end{figure}

 \begin{figure}
 \centering\includegraphics[width=\hsize]{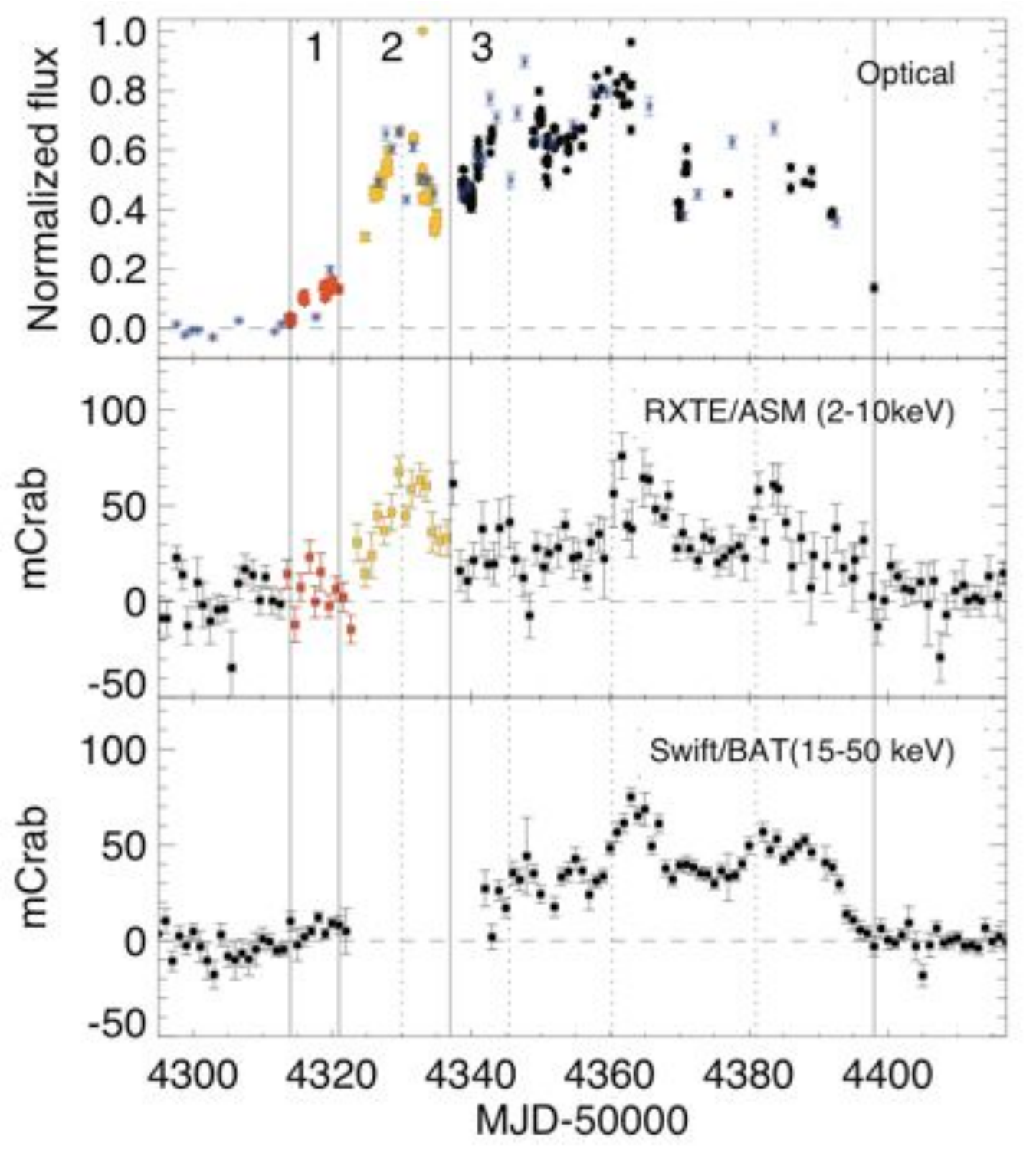}
 \caption{The optical and X-ray light curve during the 2007 outburst. The optical OGLE and
   the MOA light curves are combined, shown in the top
   panel, but the OGLE data are indicated with the dark blue symbol
   `$\ast$', and MOA are the rest.  The whole outburst is divided into
   three stages: (1) the initial rise, (2) the first peak and (3) the
   rebrightening stage.}
 \label{Fig.3}
\end{figure}

 \begin{figure}
 \centering\includegraphics[width=\hsize]{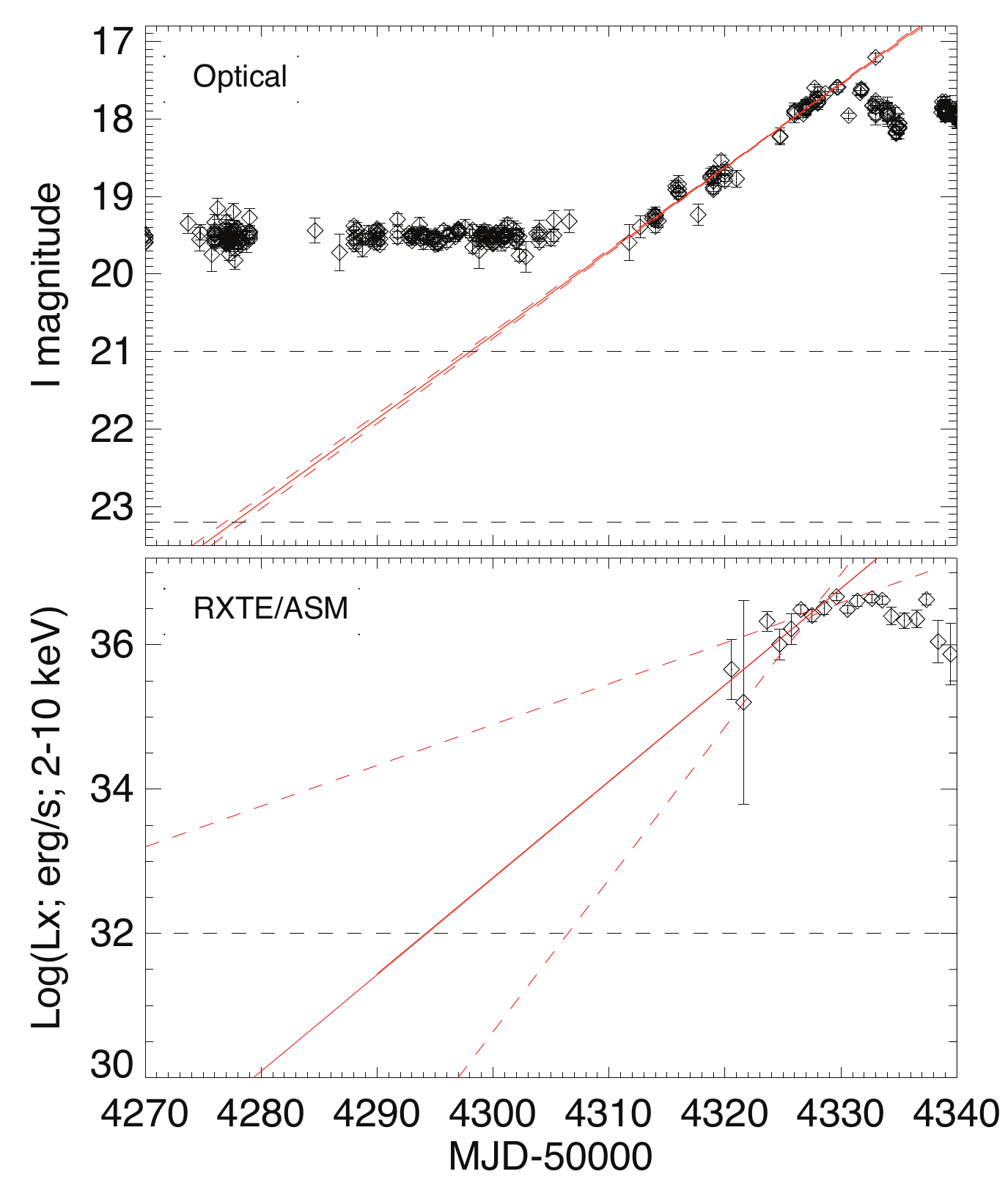}
 \caption{The best-fit power law to the points during the rise of the
   outburst. The dashed lines indicate the one sigma errors of the fits. The optical flux started to rise at MJD = $54288^{+10}_{-10}$ with $23.2 > I > 21$ magnitude in quiescence. The X-rat fit gave an X-ray initial rise at at MJD $\sim
   54293^{+12}_{-35}$. The starts points of X-ray and optical flux are consistent with each other within one-sigma error}
 \label{Fig.4}
\end{figure}

 \begin{figure}
 \centering\includegraphics[width=\hsize]{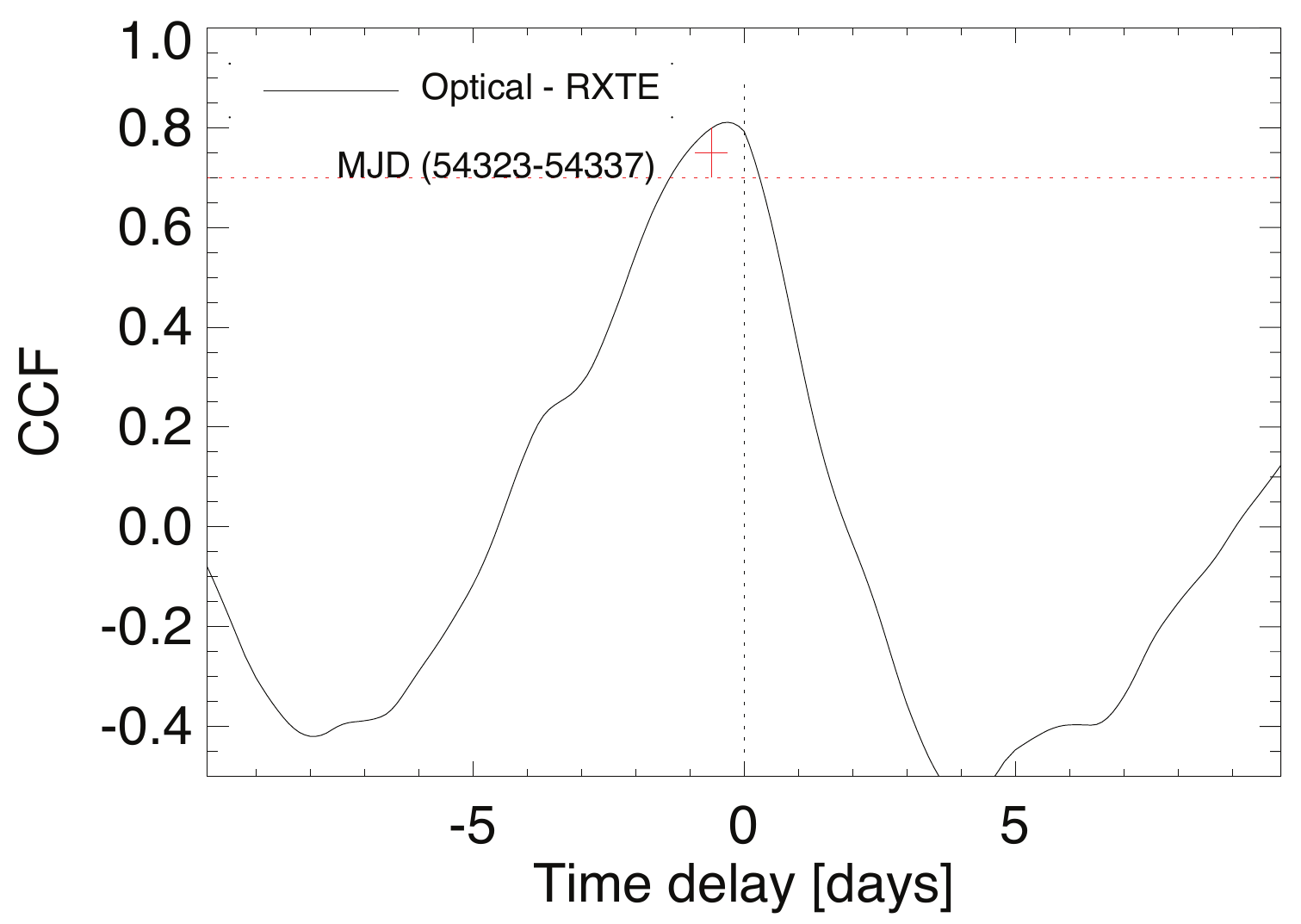}
 \caption{The CCF function of the optical and {\it RXTE} light
   curve during the first peak phase (MJD $\sim$ 54323-54337). The red
   dotted line at CCF=0.7 shows the $3\sigma$ confidence level of
   the CCF based on the observed uncertainties in the data. The red symbol
   shows the CCF area centroid above $3\sigma$ level of confidence.}
 \label{Fig.5}
\end{figure}

 \begin{figure}
 \centering\includegraphics[width=\hsize]{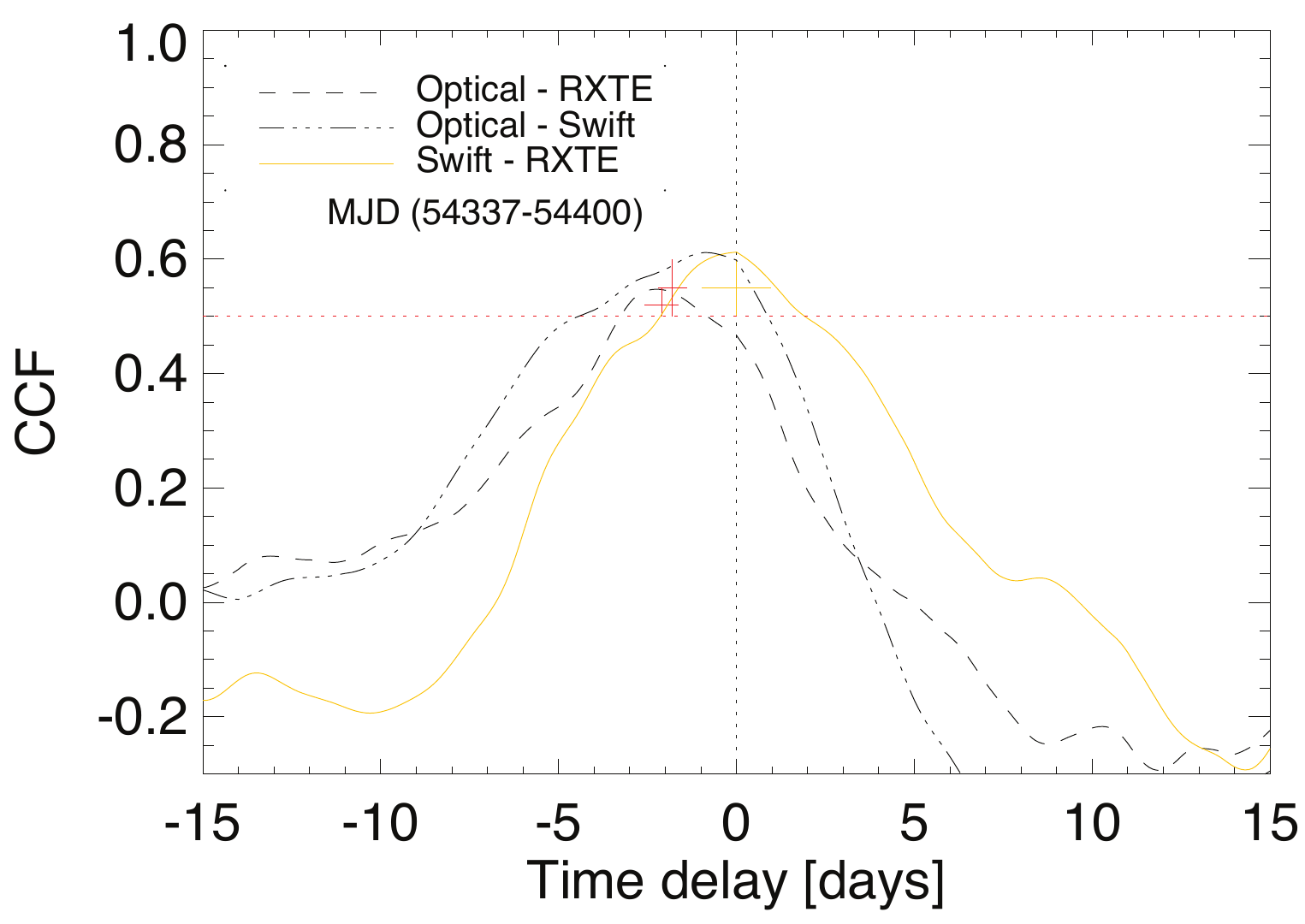}
 \caption{The CCF function of the optical and {\it RXTE}/{\it Swift}
   light curve during the rebrightening stage (MJD $\sim$
   54337-54400). The red dotted line at CCF=0.5 shows the $3\sigma$
   level of confidence. The red dashed and red dot-dashed short line
   show the CCF area centroid of optical/{\it RXTE} and optical/{\it
     Swift} separately. The yellow solid shows the CCF area centroid
   of {\it Swift}/{\it RXTE}. }
 \label{Fig.6}
\end{figure}

 \begin{figure}
 \centering\includegraphics{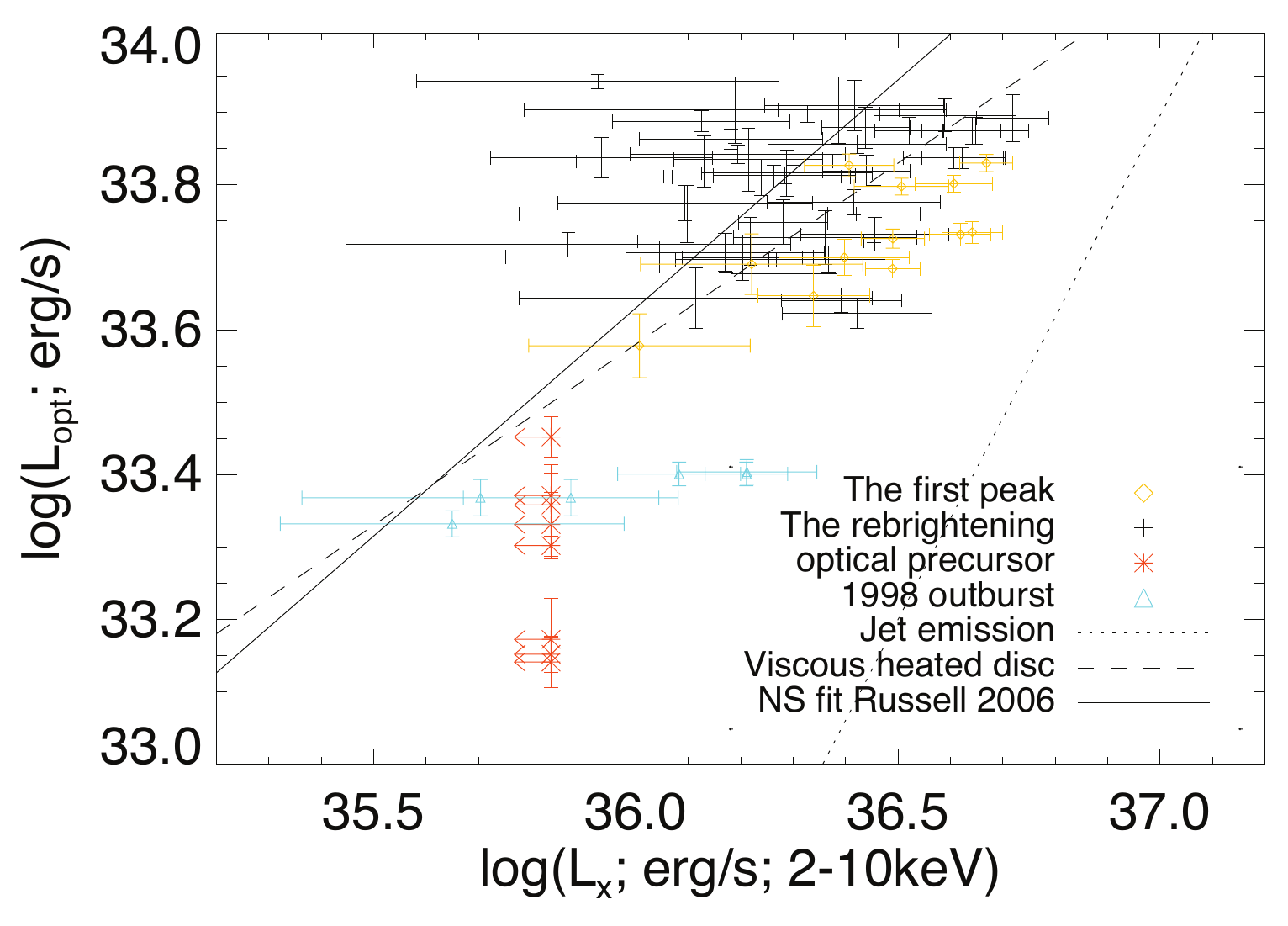}
 \caption{The plot for the quasi-simultaneous optical and X-ray (2-10
   keV) luminosities of SAX J1810.8-2609. The three lines indicate the predictions of different models. Dotted line: jet emission model, dash line:  viscously heated disk emission model, solid line: the best fit line in Russell et al. (2006) which indicates the contribution from X-ray reprocessing model.}
 \label{fig.7}
\end{figure}



\end{document}